\documentclass[fleqn,usenatbib]{mnras}

\usepackage[T1]{fontenc}
\usepackage{ae,aecompl}

\usepackage{graphicx}	
\usepackage{amsmath}	
\usepackage{amssymb}	
\usepackage{xspace}

\setcitestyle{notesep={}}

\newcommand{\dcc}{LIGO-P1800361-v5\xspace}
\newcommand{\eto}{\ensuremath{\mathrm{e}^}}
\newcommand{\tstart}{t_\mathrm{s}}
\newcommand{\fstart}{f_\mathrm{s}}
\newcommand{\vp}{vP-DV\xspace}
\newcommand{\frot}{f_\mathrm{rot}}
\newcommand{\fgw}{f_\mathrm{gw}}
\newcommand{\Egw}{E_\mathrm{gw}}
\newcommand{\Msun}{M_\odot}
\newcommand{\Iz}{I}
\newcommand{\Erot}{E_\mathrm{rot}}
\newcommand{\Sn}{S_\mathrm{n}}
\newcommand{\snr}{\rho}
\newcommand{\snropt}{\snr_\mathrm{opt}}
\newcommand{\pdet}{p_\mathrm{det}}
\newcommand{\pfa}{p_\mathrm{FA}}
\newcommand{\pfd}{p_\mathrm{FD}}
\newcommand{\thresh}{\rho_\mathrm{thr}}
\newcommand{\Tc}{T_\mathrm{c}}
\newcommand{\pyCBC}{\texttt{pyCBC}\xspace}

\title[unlikely `extended emission` from GW170817]
{Matched-filter study and energy budget suggest no detectable gravitational-wave `extended emission' from GW170817}
\author[M.~Oliver et al.]{Miquel Oliver$^1$\thanks{E-mail: miquel.oliver@ligo.org},
David Keitel$^{2,3}$\thanks{E-mail: david.keitel@ligo.org},
Andrew L. Miller$^{4,5}$,
Hector Estelles$^1$,
\newauthor{Alicia M. Sintes$^1$}
\\
$^{1}$Departament de F\'isica, Universitat de les Illes Balears, IAC3--IEEC, Cra. Valldemossa Km. 7.5, E-07122 Palma de Mallorca, Spain\\
$^{2}$University of Glasgow, School of Physics and Astronomy, Kelvin Building, Glasgow G12 8QQ, Scotland, United Kingdom\\
$^{2}$University of Portsmouth, Institute of Cosmology and Gravitation, Portsmouth PO1 3FX, United Kingdom\\
$^{4}$Universit\`a di Roma La Sapienza and INFN, Sezione di Roma, I-00185 Rome, Italy;\\
$^{5}$University of Florida, Gainesville, FL 32611, USA
\vspace{-\baselineskip}
}

\date{\dcc [draft version: 11 February 2019]} 

\pubyear{2019}

\begin{document}

\label{firstpage}
\pagerange{\pageref{firstpage}--\pageref{lastpage}}
\maketitle

\begin{abstract}
\cite{vanPutten:2018abw} have reported a possible detection
of gravitational-wave `extended emission' from a neutron star remnant of GW170817.
Starting from the time-frequency evolution and total emitted energy of their reported candidate,
we show that such an emission is not compatible with
the current understanding of neutron stars.
We explore the additional required physical assumptions to make a full waveform model, 
for example, taking the optimistic emission from a spining-down neutron star with fixed quadrupolar deformation,
and study whether even an ideal single-template matched-filter analysis
could detect an ideal, fully phase-coherent signal.
We find that even in the most optimistic case
an increase in energy and extreme parameters would be required
for a confident detection
with LIGO sensitivity as of 2018-08-17.
The argument also holds for other waveform models
following a similar time-frequency track
and overall energy budget.
Single-template matched filtering on the LIGO data around GW170817,
and on data with added simulated signals,
verifies the expected sensitivity scaling
and the overall statistical expectation.
\end{abstract}




\section{Introduction}
\label{sec:intro}

GW170817~\citep{gw170817FIRST} was the first binary neutron star coalescence
observed by the Advanced LIGO \citep{TheLIGOScientific:2014jea} and Virgo \citep{TheVirgo:2014hva} detectors
and the first gravitational-wave (GW) event with multi-messenger counterpart observations \citep{GBM:2017lvd}.
The merger remnant remained undetermined,
and several LIGO-Virgo (LVC) searches \citep{Abbott:2017dke,Abbott:2018hgk}
have found no evidence of a post-merger signal from the location of GW170817.
For various emission mechanisms, it was estimated that a signal would have needed to be unphysically energetic
to be detectable with current detector sensitivity and the deployed analysis methods.

Meanwhile, \cite{vanPutten:2018abw} (hereafter: \vp)
have reported a putative detection of GW `extended emission' lasting for several seconds after the merger.
No detailed physical model for this emission was provided,
but they attribute it to the spin-down of a remnant neutron star (NS)
with exponentially decaying rotation frequency.
Exploring possible amplitude evolutions,
a signal with the reported properties
would be very difficult to explain with conventional NS physics.
Even under optimistic assumptions and in an ideal single-template matched-filter analysis,
much more extreme parameters and an increased energy budget
would be required to detect such signals.

In Sec.~\ref{sec2} we first summarize the
signal candidate reported by \vp.
While there is no known exact physical model for the \vp candidate,
we make a first approach in using a conventional NS spindown model to
construct a representative GW template waveform,
and discuss the corresponding energy budget
and physical constraints.
We then calculate the optimal signal-to-noise ratio for
this template in Sec.~\ref{omf},
finding that,
even under optimistic assumptions,
more energy
would be required for detectable signals.
To verify these results, in Sec.~\ref{sec:search-methods} we briefly describe
a single-template matched-filter analysis on open LIGO data
for a signal with the \vp best-fit parameters,
and its null result.
We repeat this analysis on data with added simulated signals,
reproducing the optimal-SNR estimate of required GW energy for a detectable signal.
We conclude in Sec.~\ref{concl}.
Since our main waveform model was by necessity an ad-hoc choice,
the appendices include some checks of alternative waveform models with different amplitude evolution,
which are briefly summarized in the appropriate sections of the main paper,
generally supporting the results obtained for the reference model.

\section{Signal model and energy budget}
\label{sec2}

\vp have performed an analysis of GW data around GW170817
with a pipeline previously described and used in different contexts~\citep{vanPutten:2014lja,VanPutten:2016pif}.
It is a semi-coherent method,
similar in that respect e.g. to the LVC methods described in \citet{Miller:2018rbg,Oliver:2019ksl}.
But unlike these, it does not work with specific model-based template waveforms.
The single-detector data is first filtered with a bank of generic short time-symmetric templates.
Candidates are then identified through edge detection on merged multi-detector outlier spectrograms.

They report a GW signal candidate following a decaying exponential track in the time-frequency plane:
\begin{equation}
 \label{expppp}
 \fgw(t) = (\fstart- f_0) \, e^{-(t-\tstart)/\tau} + f_0 \quad \text{for} \; t>\tstart \,,
\end{equation}
where $\fstart$ is the starting frequency of the signal,
$f_0$ is the frequency that the signal asympotically approaches,
$\tau$ is a decay time scale constant,
$t$ is time,
and $\tstart$ is the reference time for $\fstart$.
The best-fit values are given as
\mbox{$\fstart=650$\,Hz},
\mbox{$f_0=98$\,Hz},
\mbox{$\tstart=0.67$\,s}
after the merger~\citep[at nominal coalescence time of $\Tc=1187008882.43$,][]{gw170817FIRST},
and \mbox{$\tau=3.01 \pm 0.2$\,s}.
The emitted GW energy in this signal is quoted as
\mbox{$\Egw\simeq0.002\,\Msun c^2$},
where $\Msun$ is a solar mass and $c$ is the speed of light.

For comparison, \citet{Abbott:2017dke}
performed unmodelled searches for short ($\lesssim 1$\,s)
and intermediate-duration ($\lesssim 500$\,s) signals.
For simulated waveforms of various types and durations,
they were sensitive to energy emission of \mbox{$0.6$--$19.6 \Msun c^2$}
\citep[see correction in footnote of][]{Abbott:2018hgk}.
No evidence of GW emission was found in either range.
The longer-duration search in \citet{Abbott:2018hgk}
was not aimed at the \mbox{$\lesssim10$\,s} relevant for the \vp candidate.
Hence, the \vp claim concerns a much weaker signal
than could have been found with those searches.

Turning to a physical understanding of the \vp time-frequency track,
one would conventionally expect the angular rotation frequency
\mbox{$\Omega = 2\pi\frot$}
of a NS to follow a solution of the general torque equation,
\mbox{$
\dot{\Omega} = - k \, \Omega^n \,,
$}
where the constant $k$ and braking index $n$ depend on the processes of energy loss.
For \mbox{$n>1$} the solution is a power law \citep{Shapiro1983,Palomba2001:mag,Lasky:2017hff},
yielding the GW waveform model considered for longer-duration postmerger signals in \citet{Abbott:2017dke,Abbott:2018hgk}.
While the \mbox{$n=1$} limit would give an exponential decay as in Eq.~\eqref{expppp},
the asymptotic $f_0$ term cannot be interpreted this way.
Still, the torque equation might not apply for an extremely young merger remnant,
or $\fgw$ might not be a simple multiple of $\frot$.
Hence, we will not attempt here to find any extended physical explanation for Eq.~\eqref{expppp},
but simply consider it as an ad-hoc input
and investigate its claimed detectability.

However, to construct a complete waveform model for matched filtering,
and to connect with the energy budget,
we also need the corresponding GW amplitude.
Again, no clearly defined model was suggested by \vp,
and hence we need to explore some possible assumptions.
As long as the signal is quasi-monochromatic following Eq.~\eqref{expppp},
we can describe it with a simple dimensionless timeseries $h_0(t)$,
corresponding to the amplitude envelope of the rapidly oscillating strain $h(t)$
at a detector induced by an optimally oriented source.
The emitted energy up to a time $T$ is then
\begin{equation}
 \label{eq:Egw}
 \Egw = \int\limits_{t=\tstart}^{T} \mathrm{d}t \, \frac{2\pi^2c^3}{5G} d^2 h_0^2(t) \fgw^2(t) \,,
\end{equation}
at a distance $d\approx40$\,Mpc from the source,
with $G$ being Newton's gravitational constant.
It is difficult to make any unique physically motivated choice for $h_0(t)$,
as the dynamics of a merger remnant during the first few seconds could
deviate from
expectations
for older objects.

We will first consider the conventional case of GW emission from a fixed quadrupolar deformation,
which seems to be the explanation implied by \vp.
This might not actually be realistic for a very young object,
but allows to study the detectability of a signal with
the Eq.~\eqref{expppp} frequency evolution
and given energy budget
under a specific amplitude model.
We later discuss how the results change when relaxing this amplitude assumption.
We aim to be as optimistic as possible,
assuming a perfectly orthogonal deformation with respect to the rotational axis, 
which results in a single frequency of GW emission at \mbox{$\fgw=2\frot$}, 
and the extreme case of an inclination \mbox{$\cos\iota = 1$},
which for a given $h_0(t)$ yields the strongest strain signal at the detectors.

For a rotating body with fixed quadrupolar deformation,
the GW amplitude at a distance $d$ would be \citep{Zimmermann:1979ip,Jaranowski:1998qm}
\begin{equation}
 \label{h00}
 h_0(t)=\frac{4\pi^2 G}{c^4}\frac{\epsilon\,\Iz}{d}{\fgw(t)}^2 \,.
\end{equation}
Here,
\mbox{$\epsilon=(I_\mathrm{xx}-I_\mathrm{yy})/I_\mathrm{zz}$} is the NS ellipticity
and \mbox{$\Iz=I_\mathrm{zz}$} its principal moment of inertia.
This does not require spin-down dominated by GW emission
(which instead of the exponential $\fgw(t)$ would yield a \mbox{$n=5$} power law),
but only that the NS has fixed $\epsilon$ and $\Iz$
while following the $\fgw(t)$ spin-down track,
including if that track is dominated by other energy loss processes.

Inserting $h_0(t)$ into $\Egw$
yields an integral over $\fgw^6$, which is explicitly performed in appendix~\ref{sec:appendix-energy}.
\footnote{A simple python implementation of the waveform model,
and also the energy equation,
is available at
\url{https://git.ligo.org/david-keitel/vanPuttenWaveform}.}
Due to the asymptotic $f_0$ term, this diverges for $T\rightarrow\infty$.
But \vp only claim an observable GW track for 7\,s (see A4.2 in their supplement)
and, with their best-fit parameters,
$\Egw$ changes only very slowly after 7\,s (e.g. by 0.1\% up to 20\,s);
with this in mind we fix \mbox{$T=7$\,s}.

In both $h_0$ and $\Egw$ expressions,
only the product $\epsilon\,\Iz$ appears,
and for fixed {\mbox{$\Egw=\,0.002\Msun c^2$}
we can eliminate it, yielding an expected initial strain amplitude of
\mbox{$h_0(t=\tstart)\approx2.38\times10^{-23}$}.
Studying the detectability of such a signal
will be the subject of the following sections.

But to get a better intuition about the physical constraints on this
(or any similar)
emission model for GWs from a young postmerger remnant NS,
we can also for a moment consider $\Iz$ and $\epsilon$ separately,
and compare with the available energy budget.
The energy stored in the NS's rotation is
\mbox{$\Erot = \tfrac{1}{2} I\,\Omega^2$}.
Additional energy could be extracted e.g. from the magnetic field or fallback accretion,
but most of the total energy budget should be lost through non-GW channels.
As a starting point, \mbox{$\Erot(\Omega=\pi\fstart)=\Egw$}
would yield \mbox{$\Iz\approx1.7\times 10^{38}$\,kg\,m$^2$},
between the `canonical' $1\times 10^{38}$\,kg\,m$^2$ typically assumed for isolated NSs \citep{Riles:2017evm}
and the $\approx4\times10^{38}$\,kg\,m$^2$ assumed in \cite{Abbott:2018hgk} for a heavy and rapidly rotating merger remnant.
Inserting this value into Eq.~\eqref{eq:Egw} and solving for the ellipticity
leads to a huge $\epsilon\gtrsim1.2$.
If instead $\Egw<\Erot$, then $\Iz$ will be larger and $\epsilon$ can be smaller,
but at most a factor of a few can be gained without making $\Iz$ unphysical.

To our knowledge there is no solid estimate for $\epsilon$ in a very young remnant NS.
But compared with theoretical and observational constraints of \mbox{$\epsilon\ll1$} for older objects~\citep{Cutler:2002,JohnsonMcDaniel:2012wg}
and even for quite young magnetars~\citep[e.g.][]{Palomba2001:mag,Lasky:2015olc,Ho:2016qqm,DallOsso:2018dos},
we see that for the model of a NS with constant $\epsilon\,\Iz$, this factor would need to be several orders of magnitude higher
than in those regimes to emit {\mbox{$\Egw\simeq\,0.002\Msun c^2$}
along the \vp signal track.
Extreme ellipticities would also typically require extreme magnetic fields,
which then may not even allow for the orthogonal rotator configuration required to generate strong GW emission~\citep{Ho:2016qqm,DallOsso:2018dos}.

It might be possible to circumvent this argument
in models with different $h_0(t)$,
e.g. through time-varying quadrupole amplitudes or different emission channels.
Since the physics of a newborn remnant NS are uncertain,
and since \vp have heuristically fitted the $\fgw(t)$ model to the detection candidate's time-frequency-track
without specific physical assumptions,
this is an attractive option.
In the next two sections,
we will mainly take the ad-hoc model defined by Eqs.~\eqref{expppp} and \eqref{h00} at face value,
constraining its detectability
as a function of $\Egw$.
However we also consider some representative examples of alternative models in appendix~\ref{sec:appendix-alternatives},
also including some that are very optimistic and not physically motivated,
finding no qualitative differences in our conclusions.

\section{Optimal matched-filter (non-)detectability}
\label{omf}

For known waveforms in stationary Gaussian noise,
the optimal detection strategy \citep[among linear filters:][]{Wainstein1962,Helstrom1968}
is matched filtering;
see e.g. \citet{maggiore2008:_gwbook,jarakrolak2009:_anagw} for modern textbook treatments.
While LIGO data is not fully Gaussian,
in the absence of sporadic short-duration glitches \citep[see e.g.][]{Zevin:2016qwy,Nuttall:2018xhi}
it can be approximated well as coloured Gaussian noise
\citep[with additional narrow spectral line artifacts,][]{Covas:2018oik}.
A strong glitch in the Livingston detector during the inspiral of GW170817 has already been subtracted,
as described in~\cite{gw170817FIRST},
from the released data \citep{gwosc:GW170817}.

To our knowledge, no sensitivity curves are available
for this specific signal type
with the \vp cross-correlation pipeline \citep[previously described and used in different contexts:][]{vanPutten:2014lja,VanPutten:2016pif}.
But for any specific given waveform
it cannot be more sensitive than matched filtering.
Note that the appeal of semi-coherent methods such as that of \vp is
that they can retain most of their sensitivity even if an actual signal is not fully phase-coherent.
However, here we will study the optimistic case of a signal fully coherently following Eq.~\eqref{expppp};
if such an ideal signal is not detectable by fully-coherent matched filtering,
then any practical analysis cannot be more sensitive either for signals with loss of phase coherence.

To quantify detectability, let us first define the product of a template with a data stream
which leads to the complex matched-filter output function
\citep[following the notation of][]{Allen:2005fk}
\begin{equation}
 \label{eq:matched-filter-z}
 z(\tstart) = 4 \int\limits_0^\infty \mathrm{d}f \; \frac{\tilde{h}_\mathrm{data}(f) \tilde{h}^*_\mathrm{template}(f,\tstart=0,\phi_0=0)}{\Sn(f)} \, \mathrm{e}^{2\pi\,if\tstart}
\end{equation}
where $\tilde{h}(f)$ is the Fourier transform of the $h(t)$ timeseries
and $\Sn(f)$ is the single-sided noise power spectral density (PSD) of a detector.
Provided the PSD is reasonably well estimated,
this whitening factor will take proper care of any spectral noise artifacts in the data.

Before looking at actual data,
let us consider the optimal signal-to-noise ratio (SNR) $\snropt$ for a waveform template $h(t)$.
In the frequency domain, this is given \citep{Flanagan:1997sx} by
\begin{equation}
 \label{eq:snr-opt}
 \snropt^2 = 4 \int\limits_0^\infty \mathrm{d}f \; \frac{|\tilde{h}(f)|^2}{\Sn(f)} \,.
\end{equation}
The optimal SNR corresponds to a scalar product of the template with itself and it is
a measure for the sensitivity of the detector to such a given template;
this quantity is commonly used as a normalization factor when constructing the matched-filter SNR detection statistic
\begin{equation}
 \label{eq:matched-filter-rho}
 \rho(\tstart) = \frac{|z(\tstart)|}{\snropt} \,.
\end{equation}

Taking the absolute value of the complex $z(\tstart)$
is equivalent to optimizing over an unknown phase offset $\phi_0$,
so that $\rho(\tstart)$ is a SNR timeseries for sliding the template against the data.
In contrast with coalescence searches,
we use the signal start time $\tstart$ instead of the end-time as a reference.

We use the \pyCBC matched-filtering engine \citep{pycbc1.13.0}
that was also used in one of the pipelines detecting GW170817 \citep{Usman:2015kfa,gw170817FIRST}.
The PSD estimate \citep{Cornish:2014kda,Littenberg:2014oda} is taken from \citet{Abbott:2018wiz}.\footnote{A file
with these PSDs is available for download at
\mbox{\url{https://dcc.ligo.org/public/0150/P1800061/010/GW170817_PSDs.dat}}.
We have checked that the SNRs change by no more than 5\%
when instead using a simpler \pyCBC Welch's method estimate of the PSD
from the GWOSC 2048\,s data set.}

We construct the strain $h(t)$ at a GW detector from Eqs.~\eqref{expppp} and \eqref{h00}
with the frequency-evolution parameters given by \vp,
the distance \mbox{$d=40$\,Mpc} and sky location of GW170817,
the best-case \mbox{$\cos\iota=1$},
and a factor $\epsilon\,\Iz$ matching the \mbox{$\Egw=0.002\,\Msun c^2$} budget,
using standard LALSuite \citep{lalsuite} functions
to apply the detector response.
\citep[See][ for the full equations.]{Jaranowski:1998qm}

For both LIGO detectors at Hanford (H1) and Livingston (L1),
we obtain \mbox{$\snropt\approx1.8$},
which is a rather low value, as we will see in the following.
In Gaussian noise, the squared SNR is $\chi^2_\kappa$-distributed with $\kappa=2$ degrees of freedom,
with mean of 2 and variance of 4,
which in the presence of a signal becomes a non-central $\chi^2$
with mean $2+\snropt^2$.
Thus,
a \vp type signal with the given parameters and the Eq.~\eqref{h00} amplitude model
should not be confidently detectable with aLIGO at its sensitivity at the time of GW170817,
since there is significant overlap between the pure-noise and noise+signal distributions.

For a given threshold $\thresh$,
the false-alarm probability
and false-dismissal probability
-- for a single trial, i.e. a fixed start time and single waveform template --
are given by:
\begin{eqnarray}
 \pfa(\thresh)         = \int_{\thresh}^{\infty} p(\rho|\mathrm{noise})        \, \mathrm{d}\rho \,, \\
 \pfd(\thresh,\snropt) = \int_0^{\thresh}        p(\rho|\mathrm{noise+signal}) \, \mathrm{d}\rho \,,
\end{eqnarray}
where the `noise+signal' model is evaluated at fixed $\snropt$,
and the detection probability is \mbox{$\pdet=1-\pfd$}.

Still assuming Gaussian noise,
the matched-filter SNR from Eq.~\eqref{eq:matched-filter-rho}
is a Neyman-Pearson optimal statistic: it maximizes $\pdet$ at fixed $\pfa$.
$\thresh$ is thus usually chosen at an acceptable $\pfa$ level
after taking into account the trials factor
from analysing a certain length of data
and multiple templates.
To deal with long data stretches and contamination by non-Gaussian noise artifacts,
\mbox{$\thresh=8$} is often considered \citep[e.g.][]{Aasi:2013wya};
however for targeted post-merger searches where only a short interval of time is of interest,
thresholds as low as 5 have been suggested \citep{Clark:2014wua,Clark:2015zxa}.

\begin{figure}
 \includegraphics[width=\columnwidth]{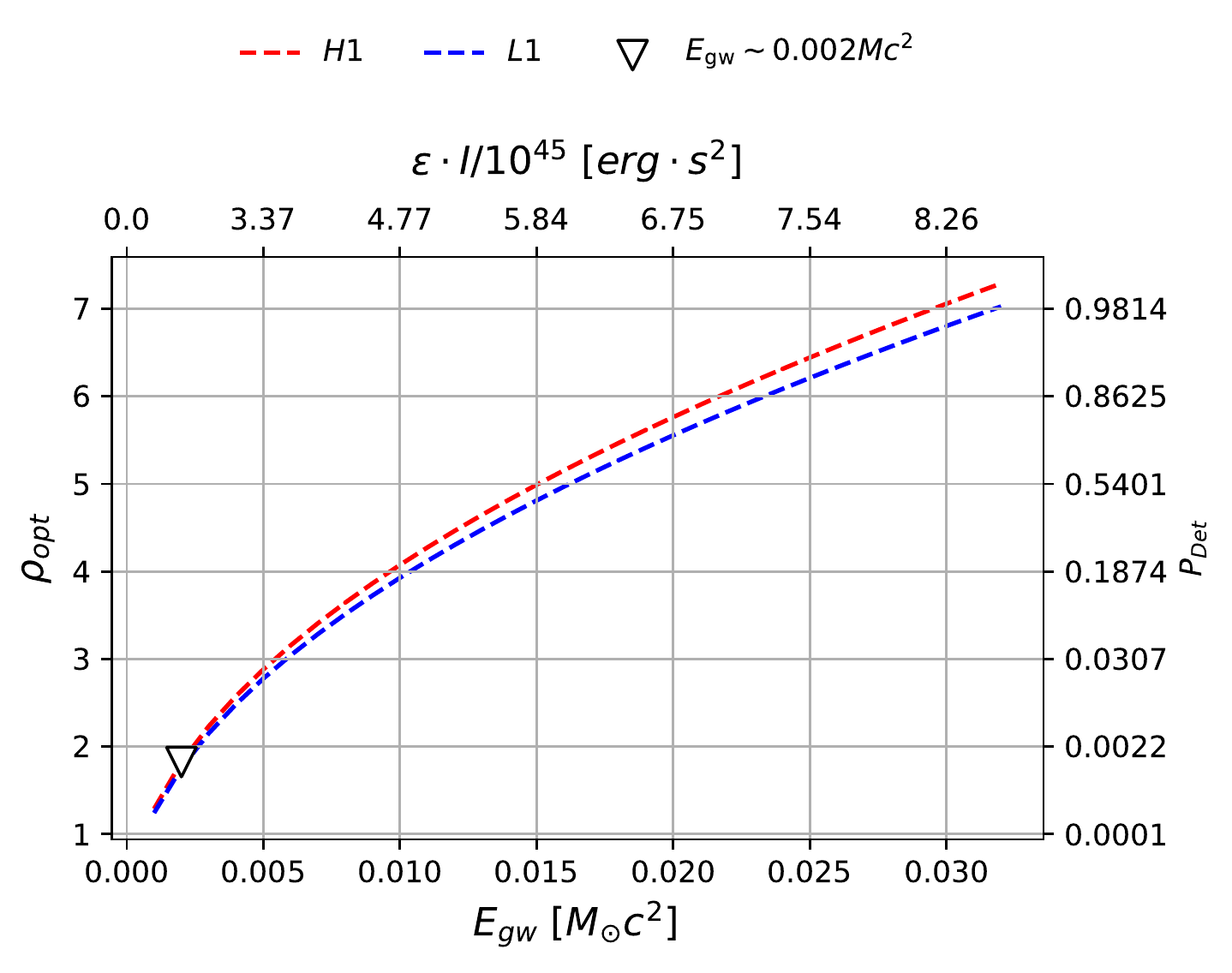}
 \vspace{-\baselineskip}
 \caption{Detectability of a \vp type signal after GW170817,
          assuming a constant-quadrupole model,
          according to the optimal SNR $\snropt$
          in each LIGO detector (H1 and L1).
          The factor $\epsilon\,I$ is adapted to scale the emitted energy $\Egw$,
          keeping all other waveform parameters fixed.
          The detection probability $\pdet$ is evaluated at a SNR threshold of 5.
          \vspace{-\baselineskip}
         }
 \label{fig:snr-opt-Egw}
\end{figure}

The optimal SNR is proportional to $1/d$, $\sqrt{\Egw}$ or $\epsilon\,\Iz$ respectively.
For illustration, Fig.~\ref{fig:snr-opt-Egw} shows the scaling with both $\Egw$ and $\epsilon\,\Iz$
assuming the fixed-quadrupole amplitude model from Eq.~\eqref{h00},
as well as the corresponding $\pdet$ at a nominal threshold of 5.
We see that a much higher emitted energy over the proposed exponential track
would be needed to make a signal confidently detectable.
In addition, Fig.~\ref{fig:PFD-PFA-RHOopt} illustrates the general relations between
$\pdet$, $\pfa$, $\snropt$ and $\thresh$
for the $\chi^2_2$ distribution,
which do not depend on the waveform model.
Choosing \mbox{$\thresh=5$} yields a single-trial \mbox{$\pfa\approx4\times10^{-6}$},
which e.g. for data sampled at 4096\,Hz corresponds to
about one false alarm per minute,
or $\approx0.025$ expected noise events above threshold
for the 1.7\,s window between GW170817 and GRB170817A \citep{Monitor:2017mdv}
suggested as a reference duration by \vp.

\begin{figure}
 \includegraphics[width=\columnwidth]{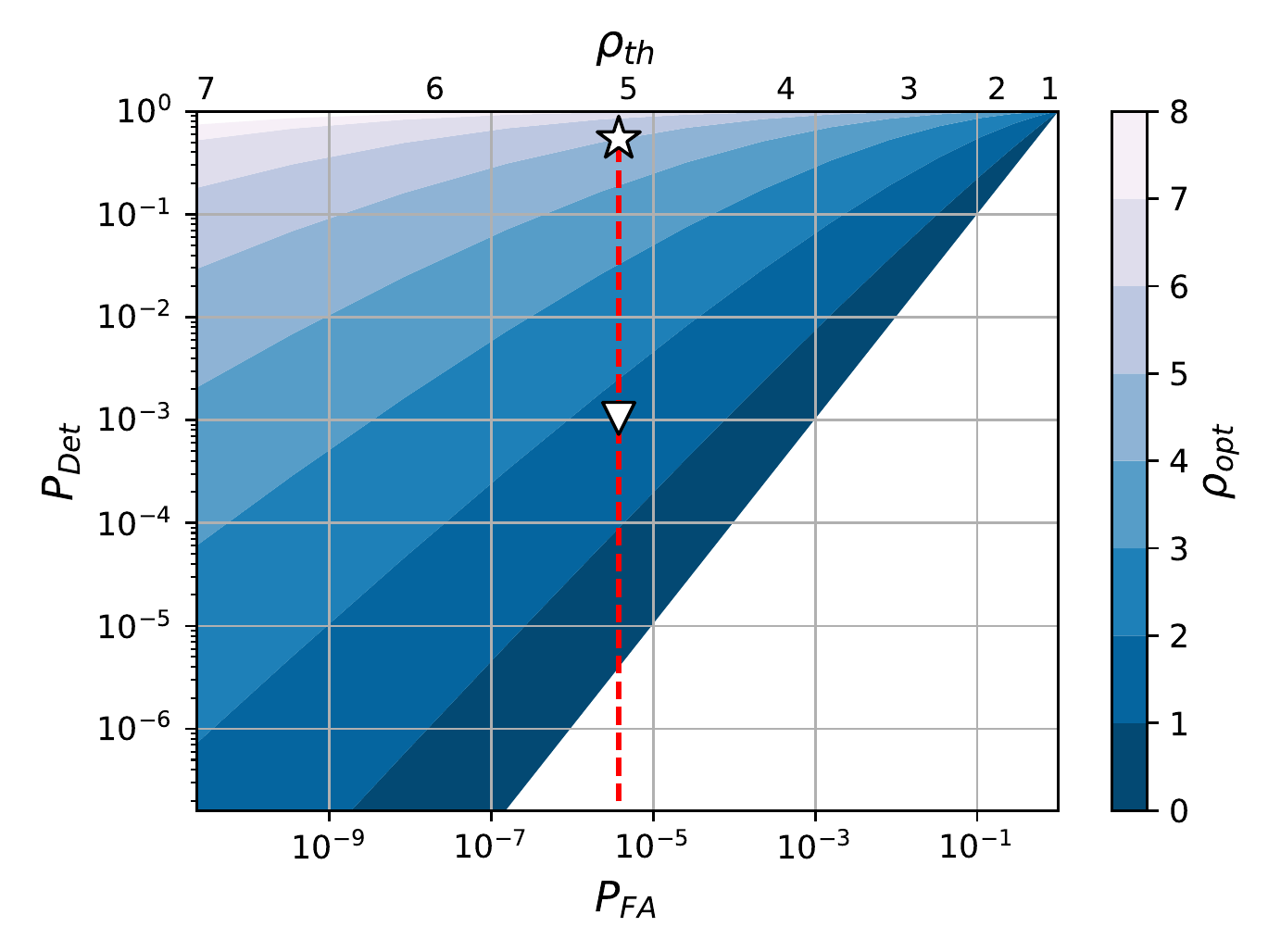}
 \vspace{-\baselineskip}
 \caption{
   Single-trial detection and false-alarm probabilities
   for the single-detector matched-filter SNR.
   Each $\pfa$ corresponds to a fixed threshold $\thresh$
   (given by the two horizontal axes)
   while $\pdet$ (vertical axis) also depends on
   the optimal SNR $\snropt$ of the signal population (colour scale).
   For an example threshold \mbox{$\thresh=5$},
   the star indicates the $(\pfa,\pdet)$ operating point
   for signals with $\snropt$ near threshold,
   while the triangle corresponds to the \mbox{$\snropt\approx1.8$}
   we obtain for the best-fit \vp parameters
   and \mbox{$\Egw=0.002\,\Msun c^2$}.
   \vspace{-\baselineskip}
   }
 \label{fig:PFD-PFA-RHOopt}
\end{figure}

If we consider this $\pfa$ level acceptable for that narrow time window of interest,
\mbox{$\snropt\approx1.8$} for the suggested \vp signal parameters and energy
yields a negligible \mbox{$\pdet(\thresh=5)\sim10^{-3}$}.
Single-detector $\pdet$ of 50\% and 90\% would require 8 and 12 times, respectively,
higher energy content of the emitted signal than suggested by \vp,
with the associated problems of making the supposed NS spindown model work
becoming correspondingly more grave.
Note again that here we have assumed optimal orientation,
\mbox{$\cos\iota=1$},
making these estimates rather conservative.

\begin{figure*}
 \includegraphics[width=\textwidth]{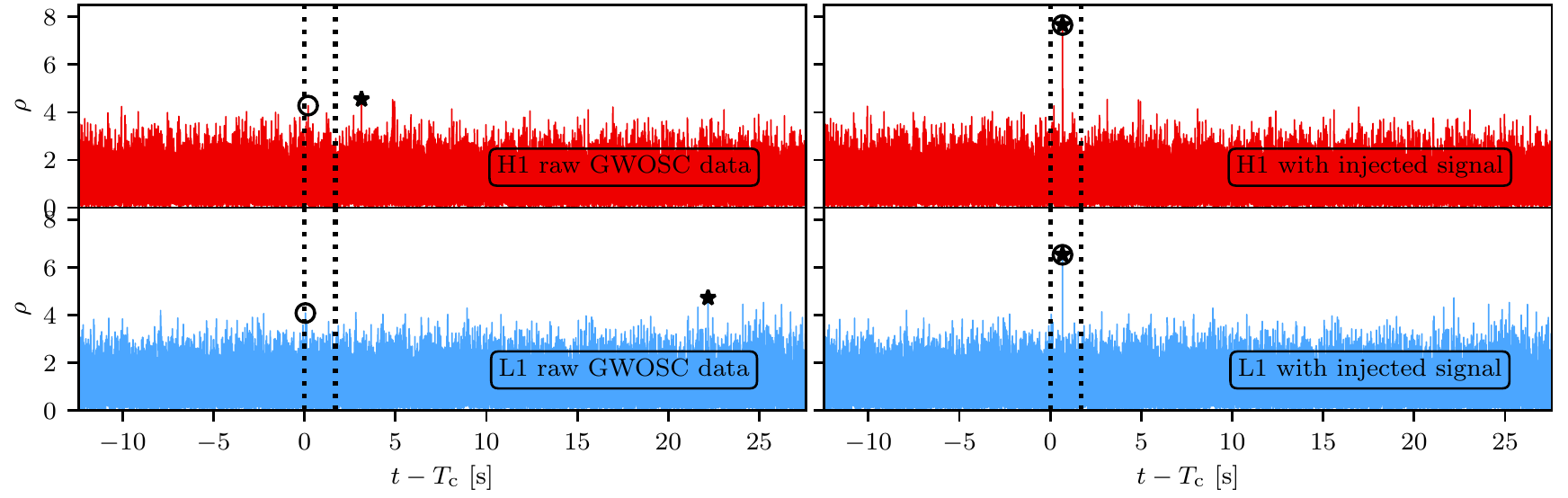}
 \vspace{-\baselineskip}
 \caption{Frequency-domain matched-filter SNR time series $\snr(\tstart)$
          obtained with \pyCBC
          on GWOSC LIGO data around the time of GW170817
          for the best-fit waveform parameters reported by \vp
          and a constant-quadrupole model.
          Stars mark the loudest outlier in a whole analysis window,
          and open circles mark the loudest candidate in a 1.7\,s window after merger time
          (\mbox{$\Tc=1187008882.43$} GPS seconds).
          First column: raw data with no significant coincident peak
          (loudest peaks of
          \mbox{$\rho_\mathrm{H1}=4.28$}
          and
          \mbox{$\rho_\mathrm{L1}=4.08$}
          separated by 0.15\,s).
          Second column: data with an injected signal matching the \vp best-fit parameters
          but a factor $\approx16$ higher emitted energy,
          with consistent peaks recovered in both detectors at \mbox{$\Tc+0.67\,s$}.
         }
 \label{fig:snr-fd-gwosc}
\end{figure*}

The combination of both LIGO detectors cannot improve the situation sufficiently either.
While e.g. the standard \pyCBC search
\citep{Canton:2014ena,Usman:2015kfa}
uses coincidences of single-detector peaks,
in principle the optimal $\pdet$ would be obtained by
coherent combination of the detector data~\citep[e.g.][]{Bose:1999pj,Cutler:2005hc,Harry:2010fr}.
The optimal approach yields an expected sensitivity improvement of $\sqrt{2}$ in amplitude,
which is insufficient to bring the \vp signal into a confidently detectable regime.
Furthermore, to be robust on real (glitchy) data this approach needs to be augmented
by additional coincidence criteria~\citep[e.g.][]{Veitch:2009hd,Keitel:2013wga,Isi:2018vst},
so this factor can be considered as an upper limit of achievable improvement.
Meanwhile, in an actual search a higher threshold would also be required
to account for the additional trials factor from searching many templates with different parameters,
further decreasing $\pdet$.

In summary, even under the most optimistic assumptions we find
that the proposed signal,
with our reference amplitude model,
only produces a low $\snropt$.
Even with lenient thresholds
and not accounting for the additional trials factor from multiple search templates,
this $\snropt$ results in a negligible detection probability.
A confident detection would require a large increase in emitted GW energy.

In appendix~\ref{sec:appendix-alternatives} we discuss optimal SNRs for alternative amplitude evolutions.
GWs from the r-mode emission channel produce slightly lower SNRs than the mass quadrupole model.
Highly optimistic ad-hoc models with $h_0(t)$ either constant or following the detector PSD,
which result in less energy emitted early on when $\fgw(t)$ corresponds to low detector sensitivity
and more emitted in the `bucket' region of best sensitivity,
can still only bring $\snropt$ up to $\sim3$--$3.6$ and thus $\pdet(\thresh=5)$ to 3--10\%.
There is no physical motivation for those ad-hoc models,
and to make up for the decaying frequency, there would need to be a
quadrupole growing by 2 orders of magnitude over the signal duration.

\section{Practical checks on real and simulated data}
\label{sec:search-methods}

Based on the optimal SNR alone, we have argued that a higher $\Egw$ would be needed
for a detectable signal of the \vp type.
To verify this, here we will demonstrate that,
as expected,
a single-template matched filter
does not return interesting candidates on the actual post-merger detector data,
and neither when a simulated signal of the suggested energy is injected;
but we can recover injections when their strength is sufficiently increased,
as predicted in the previous section.

We apply a matched-filter analysis to
the openly available LIGO data \citep{gwosc:GW170817} around the time of GW170817,
but restrict it to
the best-fit waveform parameters reported by \vp,
essentially sliding a template of fixed shape against the data
while only varying its start time and phase.
A more computationally expensive full search
over alternative waveform templates in the same Eq.~\eqref{expppp} family,
i.e. over certain ranges in all parameters,
does not seem warranted
given the expected non-detectability inferred from energy budget considerations
and optimal SNR results.

We perform \pyCBC matched filtering over 64\,s of data around the merger time of GW170817.
The only pre-processing step is a high-pass filter with cutoff at 15\,Hz to remove strong low-frequency noise components of the LIGO data,
all other features of the noise spectrum being sufficiently addressed by whitening with the noise PSD
in the matched-filter scalar product.

\begin{figure}
 \includegraphics[width=\columnwidth]{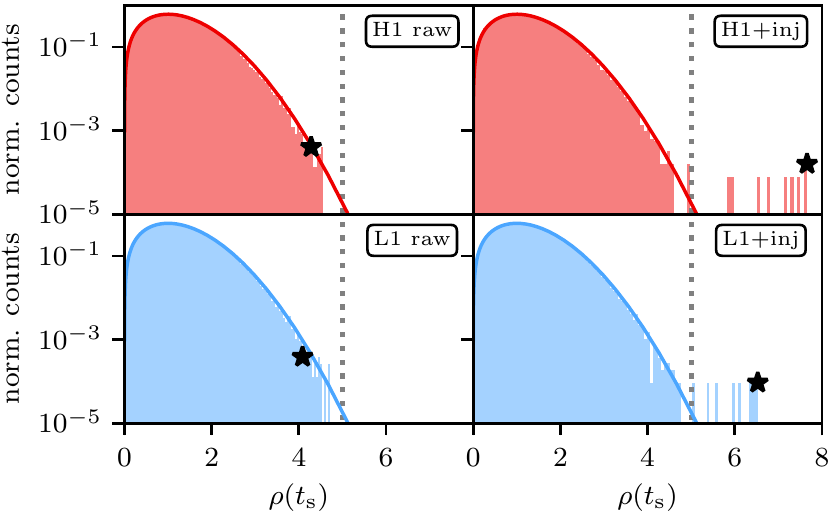}
 \vspace{-\baselineskip}
 \caption{Histograms of the SNR timeseries from Fig.~\ref{fig:snr-fd-gwosc}
          (for the best-fit waveform parameters).
          The solid line indicates the expected $\chi^2_2$ distribution in Gaussian noise,
          the black stars in each case mark the loudest candidate from a
          \mbox{$[\Tc,\Tc+1.7\,\mathrm{s}]$} window,
          and the dotted grey line corresponds to a nominal threshold of \mbox{$\rho=5$}.
         }
 \label{fig:snr-fd-hists}
\end{figure}

\begin{figure*}
 \includegraphics[width=\textwidth]{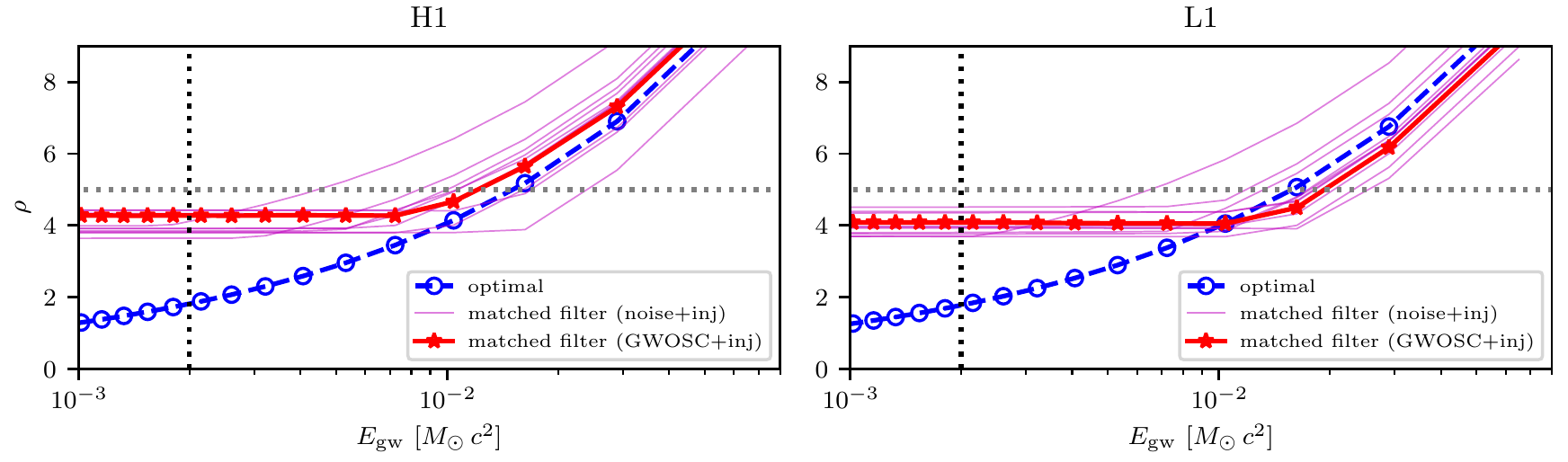}
 \vspace{-\baselineskip}
 \caption{Comparison of optimal SNRs
          and single-template matched-filter SNRs obtained on data with injected signals,
          with both injections and templates using the best-fit \vp parameters,
          as a function of changing the GW energy content $\Egw$ of the injection.
          The matched-filter SNRs are maximized over start times $\tstart$
          within the nominal 1.7\,s window between GW170817 and GRB170817A.
          The solid red line is for injections into real LIGO data around GW170817,
          and the weaker magenta lines are for the same simulated signals added to
          several realizations of coloured Gaussian noise following a smoothed PSD.
          We see that the matched-filter SNR follows the optimal SNR (dashed blue line) for strong injections,
          but already far above the energy reported by \vp
          (dotted vertical line)
          it flattens out to an injection-strength-independent level,
          which in both detectors is consistent with the spread in Gaussian noise realizations.
          A nominal $\rho=5$ threshold is also indicated (dotted horizontal line).
         }
 \label{fig:snr-fd-inj}
\end{figure*}

The matched-filter SNR does not depend on any overall amplitude normalization constant,
but does depend on the shape of $h_0(t)$.
Fig.~\ref{fig:snr-fd-gwosc}
shows the results for the model from Eq.~\eqref{h00} (constant quadrupolar deformation).
As seen in the left panels for the raw detector data,
there are abundant single-detector outliers of $\rho\approx4$,
but no particularly prominent local peaks.
In particular, \vp have emphasized the importance of their signal candidate falling
within the 1.7\,s window between GW-inferred merger time and the following GRB signal.
The loudest matched-filter SNR peaks within this window
reach only
\mbox{$\rho_\mathrm{H1}=4.28$}
and
\mbox{$\rho_\mathrm{L1}=4.08$}.
(Corresponding to single-trial $\pfa$ of $10^{-4}$ and $2.4\times10^{-4}$
from a $\chi^2_2$ distribution,
i.e. about one expected outlier of this strength or higher
from pure Gaussian noise for the $4096\times1.7$ trials within the window.)
In each detector there are several louder peaks within tens of seconds around this window,
and even those are fully compatible with Gaussian noise expectations,
see Fig.~\ref{fig:snr-fd-hists}.
Furthermore, the loudest single-detector peaks do not line up with each other
and there is no coincident peak with both single-detector SNRs above 4.

Injecting
simulated signals following the \vp waveform model
with varying amplitudes
and repeating the analysis,
we confirm that about an order of magnitude more total emitted GW energy would be required
for a confidently detectable signal under this model.
See the right column of Fig.~\ref{fig:snr-fd-gwosc} for an example SNR timeseries with a clearly recoverable injection
(of $\approx15$ times higher $\Egw$),
and Fig.~\ref{fig:snr-fd-inj} for the scaling of both optimal and matched-filter SNR with injected $\Egw$.
Again we note that this is a single-template analysis for a fully-coherent template
perfectly matching a fully-coherent injection,
which sets an upper limit on the sensitivity achievable with any
realistic (coherent or semi-coherent) search for
any (fully or only partially coherent) signals following the same frequency and amplitude evolution.

Furthermore, the obtained SNRs both without and with injections
remain consistent when exchanging the real LIGO data for simulated coloured Gaussian noise generated from a smoothed PSD,
demonstrating that the real data set is close enough to Gaussian
and the Gaussian-noise assumption inherent in the matched-filter calculation did not bias the results.
As an additional check,
we have compared these results with an independent time-domain matched-filtering implementation
in appendix \ref{sec:appendix-methods-tdmf},
again finding consistent results.

\section{Conclusions}
\label{concl}

We have investigated the detection candidate for GW170817 post-merger gravitational waves
reported by \citet{vanPutten:2018abw}.
Even in the best case,
i.e. an ideal matched-filter analysis
for a fully phase-coherent signal following their best-fit time-frequency evolution model,
and under additional optimistic assumptions,
an increase in energy and extreme parameters would be required
for a confident detection
under LIGO sensitivity at the time of GW170817.
By extension, any wide search like that of \vp should not be sensitive to this class of signals
at the expected energy budget
and at current detector sensitivity.
Hence, while a detection of post-merger GWs would have profound consequences,
this study suggests that no claim of
such a signal
from a neutron star remnant can be made at this point.

It could still be possible that \vp found a real signature in the detector strain data,
but that it might be an artifact of terrestrial origin.
While neither our (single-template) matched-filter analysis following the time-frequency evolution of the \vp candidate,
nor the previous generic post-merger searches \citep{Abbott:2017dke,Abbott:2018hgk},
have found any suspicious outliers,
the specific filtering implementation of the \vp pipeline
could react differently to artifacts.
More detailed characterization of their pipeline on simulated or off-source data
would be necessary to understand the candidate's provenance.
But even with optimistic model and parameter choices,
there seems to be no realistic possibility of an astrophysical origin.

\section*{Acknowledgements}

We thank LIGO and Virgo collaboration members including
G.~Ashton,
S.~Banagiri,
M.A.~Bizouard,
J.~Clark,
M.~Coughlin,
R.~Frey,
I.~Harry,
I.S.~Heng,
J.~Kanner,
A.~Kr{\'o}lak,
P.~Lasky,
A.~Lundgren,
M.~Millhouse,
C.~Palomba,
P.~Schale,
L.~Sun,
and G.~Vedovato
for many fruitful discussions.
M.O., H.E. and A.M.S. acknowledge the support of the Spanish
Agencia Estatal de Investigaci\'on
and Ministerio de Ciencia, Innovaci\'on y Universidades
grants FPA2016-76821-P, FPA2017-90687-REDC, FPA2017-90566-REDC, and FPA2015-68783-REDT,
the Vicepresidencia i Conselleria d'Innovaci\'o, Recerca i Turisme del Govern de les Illes Balears,
the European Union FEDER funds,
and the EU COST actions CA16104, CA16214 and CA17137.
During part of this work D.K. was funded through EU Horizon2020 MSCA grant 704094.
Some calculations were performed on the LIGO Caltech cluster.

\appendix
\section{GW energy integral}
\label{sec:appendix-energy}

Inserting the amplitude model from Eq.~\eqref{h00} into Eq.~\eqref{eq:Egw},
the emitted GW energy along a \vp type signal track from start time $\tstart$ up to an end time T
is obtained from the sixth-order integral of the frequency evolution from Eq.~\eqref{expppp}:
\begin{equation}
 \label{eq:Egw-appendix}
 \Egw = \int\limits_{t=\tstart}^{T} \mathrm{d}t \, \frac{32G}{5c^5} \Iz^2 \epsilon^2 \pi^6 \fgw^6(t)
      = \frac{32\pi^6G}{5c^5} \Iz^2 \epsilon^2 F(T)
\end{equation}
with
\begin{align}
 \label{eq:Egw-T-integral}
 F(T) =& \int\limits_{t=\tstart}^{T} \mathrm{d}t \, \fgw^6(t) & \nonumber \\
      =&  f_0^6\,(T-\tstart) + \tau \; (
                    & & 6             & (\eto{- (t-\tstart)/\tau} - 1) f_0^5            (f_0-\fstart)\hphantom{^1} \nonumber \\
       &            &-& \tfrac{15}{2} & (\eto{-2(t-\tstart)/\tau} - 1) f_0^4            (f_0-\fstart)^2 \nonumber \\
       &            &+& \tfrac{20}{3} & (\eto{-3(t-\tstart)/\tau} - 1) f_0^3            (f_0-\fstart)^3 \nonumber \\
       &            &-& \tfrac{15}{4} & (\eto{-4(t-\tstart)/\tau} - 1) f_0^2            (f_0-\fstart)^4 \nonumber \\
       &            &+& \tfrac{ 6}{5} & (\eto{-5(t-\tstart)/\tau} - 1) f_0              (f_0-\fstart)^5 \nonumber \\
       &            &-& \tfrac{ 1}{6} & (\eto{-6(t-\tstart)/\tau} - 1) \hphantom{f_0^0} (f_0-\fstart)^6 \nonumber \\
       &          & )
\end{align}
for $T>\tstart$.

\section{Time-Domain Matched Filtering}
\label{sec:appendix-methods-tdmf}

Since the waveform model of Eq.~\eqref{expppp} is quasi-monochromatic
(dominated by a single frequency at each time step, and varying over slower timescales than the inverse of the GW frequency)
we can equivalently compute SNRs with a simple time-domain approach.
A time-domain scalar product between two time-series is given by
\begin{equation}
 \left\langle h_1(t)\,|\,h_2(t) \right\rangle
 = 2 \sum\limits_{k=1}^{T_\mathrm{obs}/\mathrm{d}t} \mathrm{d}t \; \frac{h_1(t_k)\,h_2(t_k)}{S_\mathrm{n}(f(t_k))}
\end{equation}
In analogy with the frequency-domain case, the SNR (for a fixed template reference time) is then
\begin{equation}
 \rho = \frac{|z|}{\sigma}
      = \frac{|\left\langle h_\mathrm{data}(t)\,|\,h_\mathrm{template}(t) \right\rangle|}
             {\sqrt{\left\langle h_\mathrm{template}(t)\,|\,h_\mathrm{template}(t) \right\rangle}} \,.
\end{equation}

To deal with narrow spectral artifacts \citep[lines, see][]{Covas:2018oik}
in the LIGO data,
we could go to the frequency domain, whiten,
and transform back to the time domain.
As a more independent cross-check of the FD calculation in the previous section,
we instead choose the simpler (though potentially not optimal) method
of notching out pre-selected frequency bands around strong lines
identified from the PSD.
Many different implementations of such notch filters are possible;
here we use finite-impulse-response filters from
the \texttt{scipy.signal} package
\citep{scipy1.1.0}.
A bandpass filter in [30,750]\,Hz is also applied using the same functions.
We note that it is easily possible to erroneously obtain significantly higher SNRs
from the GWOSC data
if the notches are not strict enough,
as the exponential $\fgw(t)$ track then accumulates contributions from several strong lines
in the frequency regions it traverses.

We obtain consistent results both for the optimal SNR
and for the matched-filter SNR on real or simulated data with injections following the \vp model:
optimal SNRs agree within 1\% with \pyCBC results
and on real data with injections the notches cost about 5\% of SNR.

\section{Modified signal models}
\label{sec:appendix-alternatives}

\begin{figure}
 \includegraphics[width=\columnwidth]{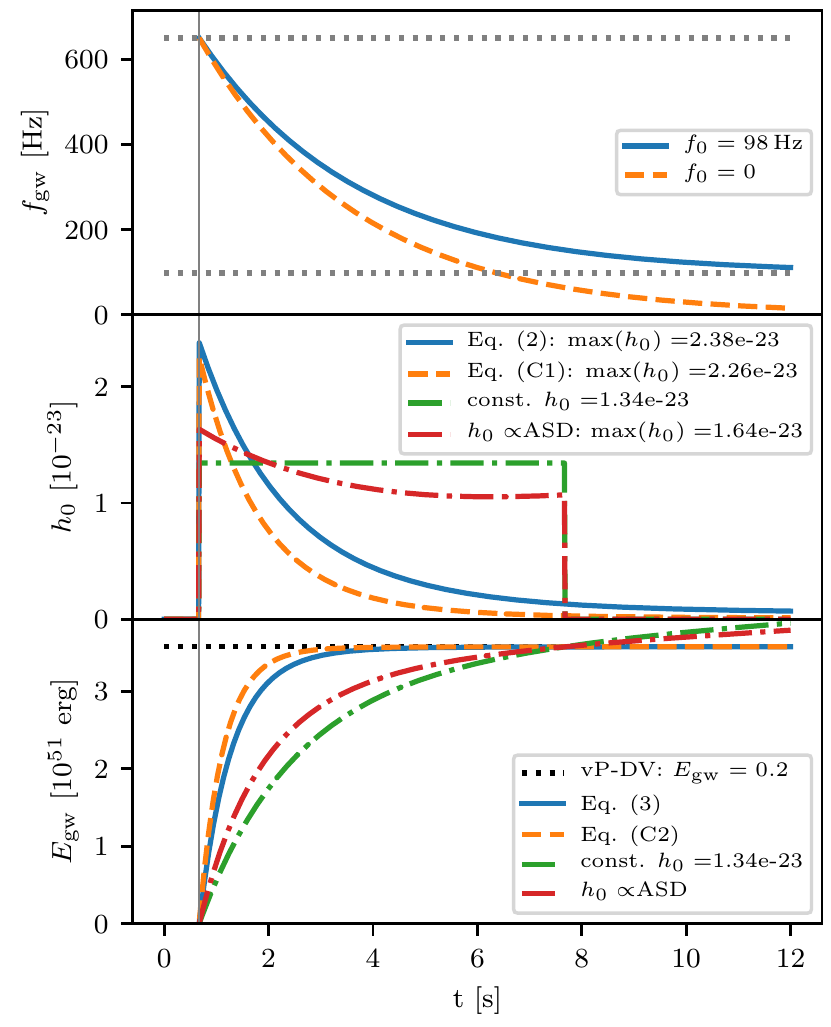}
 \vspace{-\baselineskip}
 \caption{$\fgw$, $h_0$ and $\Egw$ for the alternative models considered.
          In the top panel, $\fgw$ from Eq.~\eqref{expppp} is considered
          with and without the $f_0$ term.
          In the $h_0$ and $\Egw$ panels, \mbox{$f_0=98$\,Hz} is always assumed,
          and we show both the
          constant-quadrupole and r-mode models,
          as well as the alternative ad-hoc models
          of constant $h_0$
          or of $h_0(t)$ following a smoothed ASD at $\fgw(t)$.
         }
 \label{fig:alternative-models}
\end{figure}

In the main part of this paper,
we have assumed the $\fgw(t)$ model (Eq.~\ref{expppp}) provided by \vp
and have made the additional assumption of a constant amplitude of the quadrupole moment
(constant $\epsilon\,\Iz$)
during the NS spindown,
to get the amplitude and energy estimates
needed for matched-filtering
(Eqs.~\ref{h00} and \ref{eq:Egw-T-integral}).
These are not unique choices,
and especially given the apparent difficulties
pointed out in Sec.~\ref{sec2} to make this model physically consistent,
it seems prudent to check if our conclusion
of non-detectability still holds for reasonable modifications to the waveform model.
Let us briefly consider the following alternatives,
as also summarized in Fig.~\ref{fig:alternative-models}:
\begin{enumerate}
 \item Removing the asymptotic $f_0$ term.
 \item Considering the r-mode emission channel
       instead of GWs from a stationary deformation.
 \item An ad-hoc model with constant signal amplitude (not physically motivated).
 \item An ad-hoc model (again not physically motivated)
       where the signal amplitude tracks the detector noise PSD
       as the signal frequency decays,
       thus producing roughly constant SNR contributions
       throughout the \vp candidate's time-frequency track.
\end{enumerate}

\subsection{Setting $f_0=0$}
This makes $\fgw(t)$ decay faster,
reaching 30\,Hz\ less than 10\,s after starting at 650\,Hz.
To achieve the same emitted energy,
a slightly higher $\epsilon\Iz$ factor is required,
leading to higher initial
$h_0(t)$.
However, since the signal leaves the `bucket' region of best detector sensitivity faster,
the optimal SNRs at \mbox{$\Egw\simeq0.002\Msun c^2$}
come out about 4\% lower than for the fiducial model.

\subsection{r-mode GW emission}
A class of inertial neutron star oscillations,
r-modes can enter an unstable growing regime and become efficient GW emitters \citep{Andersson:2000mf}.
They could be an important contribution in newborn magnetars~\citep{Ho:2016qqm}.
For this example, we make the usual simplified assumption of
\mbox{$\fgw \approx \tfrac{4}{3} \frot = \tfrac{2}{3\pi}\Omega$}
\citep[for a more accurate treatment, see][]{Idrisy:2014qca}.
From Eq. (23) in~\cite{Owen:2010ng},
\begin{equation}
 \label{eq:h0-rm}
 h_0 = \sqrt{\frac{8\pi}{5}} \, \frac{G}{c^5} \, \frac{ M R^3 \tilde{J}}{d} \, \alpha \, (2\pi\fgw)^3 \,,
\end{equation}
where the NS mass $M$ and radius $R$
and the density parameter $\tilde{J}$
depend on the equation of state.
From Eq. (6) of~\citet{Ho:2016qqm},
the corresponding emitted GW energy is
\begin{align}
 \label{eq:Egw-rmodes}
 \Egw &= \int\limits_{t=\tstart}^{T} \mathrm{d}t \, \frac{96\pi}{15^2} \left(\frac{4}{3}\right)^6 \frac{G M R^4 \tilde{J}^2 I}{c^7 \tilde{I}} \alpha^2 \Omega^8(t) \\
      &= \frac{3}{25} (2\pi)^9 \frac{G M^2 R^6 \tilde{J}^2}{c^7} \alpha^2 F_\mathrm{r}(T) \nonumber
\end{align}
and inserting $\fgw(t)$ from Eq.~\eqref{expppp},
the integral
\mbox{$F_\mathrm{r}(T) = \int\limits_{t=\tstart}^{T} \mathrm{d}t \, \fgw^8(t)$}
can be evaluated in analogy with Eq.~\eqref{eq:Egw-T-integral}.

We only aim for an order-of-magnitude estimate in this section,
conservatively allowing \mbox{$M\in[2.4,3.0]\Msun$} and \mbox{$R\in[10,15]$\,km},
but simply assuming the usual \mbox{$\tilde{J}=0.01635$}.\footnote{Though
the heavy remnant of GW170817 will presumably have a quite different density structure
than the $M\approx1.4$ regime usually considered in most of the literature,
the ranges in $M$ and $R$ should be large enough to make $\tilde{J}$ not a decisive parameter.}
This yields 
\mbox{$\Egw \approx \alpha^2 \, (0.5$--$9)\times10^{48}\,\mathrm{erg} \approx \alpha^2 \, (0.2$--$5)\times10^{-6}\Msun c^2$}.
Hence an r-mode amplitude \mbox{$\alpha\in[20,100]$} would be required to match the \vp energy estimate.

Assuming equal energy,
the GW strain amplitude starts out slightly lower
and decays more quickly;
hence, the optimal SNRs are even lower than discussed in the main part of the paper,
even if, against conventional wisdom~\citep{Arras:2002dw,Bondarescu:2008qx},
\mbox{$\alpha\gg1$} would be possible.

\subsection{Constant $h_0$}
As an extreme case,
let us also make an ad-hoc model with constant $h_0$ for times
\mbox{$\tstart<t<T$},
without claiming a physical justification for it.
In this model, most of the SNR would be accumulated not at the start, but towards the end of the signal.
If the GW frequency still follows Eq.~\eqref{expppp} without an explicit cutoff $T$,
and the emitted energy follows Eq.~\eqref{eq:Egw},
it would quickly diverge.
For a cutoff \mbox{$T=\tstart+7$\,s}
(matching the track length reported in appendix A4.2 of \vp),
the nominal \mbox{$\Egw\simeq0.002\Msun c^2$} corresponds
to \mbox{$h_0\approx1.34\times10^{-23}$}
and an optimal SNR \mbox{$\snropt=3.58$} (in H1)
that yields a $\pdet\approx10\%$
at a threshold of 5.
This is not completely negligible like the $\pdet\sim10^{-3}$ obtained in Sec.~\ref{omf},
but still far from enabling confident detection.
Also note that this still corresponds to the optimal case of
\mbox{$\cos\iota=1$}.

In addition, producing a constant $h_0$ at rapidly decreasing $\fgw(t)$
requires a rapidly increasing quadrupole moment,
at the end of the track achieving a value 20 times larger than the one found in Sec.~\ref{sec2}.

\subsection{Ad-hoc $h(t)$ model for constant SNR contribution over time}
While the amplitude evolutions from Eqs.~\eqref{expppp} and \eqref{eq:h0-rm}
would lead to most SNR accumulated at the start of the signal,
another possibility to reproduce more closely the 7\,s long signal track claimed by \vp
is another ad-hoc model
where we make $h(t)$ follow the detector noise spectral density
as the signal sweeps through the band with decaying $\fgw(t)$,
i.e. we use an approximate fit
\begin{align}
 h_0(t) \propto& \sqrt{\Sn(\fgw(t))} \\
 \approx& A \, \fgw(t)^2 \left( \frac{1}{\fgw(t)^4} + \frac{0.00013125}{\fgw(t)^2} + \frac{3.1875\cdot10^{-7}}{\fgw(t)} \right) \nonumber
\end{align}
As shown in Fig.~\ref{fig:alternative-models},
this gives a bit more early and less late emission than \mbox{$h_0=\mathrm{const.}$},
but is indeed qualitatively similar.
In H1 we obtain \mbox{$\snropt=3.06$},
corresponding to \mbox{$\pdet\approx3\%$}
at a threshold of 5,
higher than our reference $h_0(t)$ model but lower than for \mbox{$h_0=$const.}
Again this would require a rapidly increasing quadrupole moment over time.

\bibliographystyle{mnras}
\bibliography{../resvp}

\bsp	
\label{lastpage}
\end{document}